# Hall effect measurements on epitaxial SmNiO$_3$ thin films and implications for antiferromagnetism

Sieu D. Ha[*1†], R. Jaramillo[*1], D. M. Silevitch[2], Frank Schoofs[1], Kian Kerman[1], John D. Baniecki[3], and Shriram Ramanathan[1]

[†]Corresponding author: sdha@seas.harvard.edu

[*]Equally contributing co-authors

[1]School of Engineering and Applied Sciences, Harvard University, Cambridge, MA 02138, USA

[2]The James Franck Institute and Department of Physics, The University of Chicago, Chicago, IL 60637, USA

[3]Fujitsu Laboratories, Atsugi 243-0197, Japan

**Abstract**

The rare-earth nickelates ($R$NiO$_3$) exhibit interesting phenomena such as unusual antiferromagnetic order at wavevector $q$ = (½, 0, ½) and a tunable insulator-metal transition that are subjects of active research. Here we present temperature-dependent transport measurements of the resistivity, magnetoresistance, Seebeck coefficient, and Hall coefficient ($R_H$) of epitaxial SmNiO$_3$ thin films with varying oxygen stoichiometry. We find that from room temperature through the high temperature insulator-metal transition, the Hall coefficient is hole-like and the Seebeck coefficient is electron-like. At low temperature the Néel transition induces a crossover in the sign of $R_H$ to electron-like, similar to the effects of spin density wave formation in metallic systems but here arising in an insulating phase ~200 K below the insulator-metal transition. We propose that antiferromagnetism can be stabilized by bandstructure even in insulating phases of correlated oxides, such as $R$NiO$_3$, that fall between the limits of strong and weak electron correlation.

## I. INTRODUCTION

Correlated-electron oxides are of intensive fundamental and applied interest due to properties such as colossal magnetoresistance, insulator-metal phase transitions, and emergent electron interactions in superlattices. However, many gaps remain in our understanding of the physics of these systems. In particular, phase transitions in nearly-itinerant correlated electron systems often confound established models that apply in the limits of weak or strong electronic correlations. For example, in the colossal magnetoresistive manganites the insulator-metal transition is characterized by strong coupling between the spin, charge, orbital, and lattice degrees of freedom, but the charge ordered state bears important signatures of a weakly coupled density wave.[1] Similarly, the rare-earth nickelates ($R$NiO$_3$) feature nearly-itinerant electrons with strongly coupled degrees of freedom, significant electron-phonon coupling and small polaron conductivity, and spectroscopy reveals the importance of the large electron correlation energy at the insulator-metal transition.[2,3] However, recent results on superlattices demonstrate that LaNiO$_3$ exhibits signatures of a weakly coupled spin density wave (SDW).[4] The nickelates can be tuned from Fermi liquids to strongly renormalized bad metals to antiferromagnetic insulators through epitaxial strain and/or chemical substitution,[3,5-7] and are therefore an important experimental platform for evaluating theories that bridge the limits of strong and weak electron correlation.

The $R$NiO$_3$ phase diagram (Fig. 1a) is controlled by the radius ($r$) of the $R^{3+}$ ion, which controls the tilts of the (NiO$_6$)$^{3-}$ octahedra, which in turn control the electronic structure.[7,8] LaNiO$_3$ is a paramagnetic metal (PM) at all temperatures, but for heavier/smaller $R^{3+}$ the ground state is an antiferromagnetic insulator (AFI). For PrNiO$_3$ and NdNiO$_3$ the Néel and insulator-metal transitions occur at the same temperature, and $T_N$ (=$T_{IM}$) increases with $R$ atomic number. However, as $R$ becomes heavier ($r$ decreases) the transition temperatures separate, a paramagnetic insulator (PI) phase emerges in the range $T_N < T < T_{IM}$, and $T_N$ decreases. Contrasting theories have been proposed to explain the magnetic and electronic phase transitions in $R$NiO$_3$, but they have limited success in explaining the full phase diagram.[9,10] Within a generic model for antiferromagnetic insulators a non-monotonic dependence of $T_N$ on the tuning parameter is associated with a crossover from localized to itinerant electrons.[11] For the Hubbard model with intra-site electron-electron interaction energy $U$ and electronic bandwidth $W$, the limit $U/W << 1$ (weak correlations) corresponds to a metal with an SDW instability and $T_N$ increases with $U/W$. In the opposite limit, $U/W >> 1$ (strong correlations), the system is a magnetic insulator described by the Heisenberg hamiltonian, and $T_N$ decreases with $U/W$. The maximum in $T_N$ therefore marks the point in the phase diagram that is most challenging to describe within the conceptual framework of either weak or strong coupling.[11,12] For $R$NiO$_3$ this maximum is realized in SmNiO$_3$.

Here we present Hall coefficient ($R_H$) measurements of SmNiO$_3$ thin films with varying oxygen content over a wide (30 – 400 K) temperature range through both Néel and insulator-metal phase transitions. There are few reports of Hall coefficient measurements on $R$NiO$_3$, and we are not aware of any for temperatures appreciably lower than $T_{IM}$.[13-15] We find that $R_H$ is hole-like in the metallic phase and it increases as temperature is lowered into the PI phase. However, $R_H$ unexpectedly changes sign to electron-like just below $T_N$. By varying $T_N$ via oxygen stoichiometry we provide evidence that the crossover in the sign of $R_H$ is connected to the onset of antiferromagnetism. We discuss the possible impact on our results of Hall coefficient anomalies arising from small polaron transport. We propose a mechanism to explain the crossover in $R_H$ that is akin to SDW formation, which is principally associated with metals but here emerges in an insulating phase. Finally, we use the case of SmNiO$_3$ to illustrate connections between concepts developed to treat the limits of electron localization (oxide physics, superexchange magnetism) and delocalization (metals physics, band magnetism). These connections suggest a flexible conceptual framework with which to understand antiferromagnetic order in the nickelates and other material systems that fall between the limits of strong and weak electron coupling.[6,16-18]

## II.  EXPERIMENTAL METHODS

$R$NiO$_3$ are thermodynamically unfavorable at typical oxide growth temperatures.[7] Most samples are actually $R$NiO$_{3-\delta}$ ($\delta > 0$), and the effects of non-zero $\delta$ can be significant.[19,20] For this work we grew SmNiO$_3$ thin films with varying oxygen content onto single crystal LaAlO$_3$ (001) substrates (MTI Corporation) by RF magnetron sputtering in relatively high background pressure from a stoichiometric target (ACI Alloys).[21] Sputtering conditions were 80/20 sccm Ar/O$_2$ gas flow ratio, 650 °C substrate temperature, and 200 W RF plasma power. Films were cooled in an ambient environment and no post-deposition annealing was performed. We controlled the oxygen stoichiometry by varying the high sputtering background pressure, as demonstrated in Ref. 21. The three samples studied in this work are labeled SNO1, SNO2, and SNO3, with growth pressures of 370, 360, and 260 mTorr and thicknesses of 15.5 ± 0.1, 18.8 ± 0.1, and 19.4 ± 0.1 nm, respectively. Growth rates were in the range of 3-5 nm/hr. Statistically meaningful values of $\delta$ could not be determined, but SNO1 is the most stoichiometric film (smallest $\delta$) and SNO3 is the least stoichiometric film (largest $\delta$). We characterized thin film crystal structure by X-ray diffraction (XRD) using a 4-circle Bruker D8 Discover diffractometer with a Göbel mirror, and we measured film thickness using X-ray reflectivity. We measured morphology by atomic force microscopy (AFM) using an Asylum MFP-3D system.

For Hall effect measurements, we patterned our films into bars (channel dimensions 400 x 2000 μm$^2$) using dilute HCl as an etchant. We measured $R_H$ in a cryostat (PPMS, Quantum De-sign) with cyclical magnetic field sweeps between ±6 T at each temperature. In order to measure $R_H$ over a wide temperature range we paid particular attention to heat sinking and thermal stabilization. The large temperature coefficient of resistivity and large overall resistivity in $R$NiO$_3$ may have limited Hall effect studies to date. The samples were heat sunk through the LaAlO$_3$ substrate, which was glued with silver epoxy to a sapphire support plate, which was in turn varnished to the cryostat cold plate. This thermal anchoring through the substrate was necessary to obtain reliable data despite the fact that the sample space in the cryostat is filled with helium exchange gas. We measured the Hall and longitudinal resistances $R_{xy}$ and $R_{xx}$ simultaneously. We averaged $R_{xy}$ between consecutive forward and backward field sweeps to account for temperature drift, and we removed the contribution of magnetoresistance to the Hall voltage by subtracting the scaled $R_{xx}$, i.e. $R'_{xy}(H) = R_{xy}(H) - (R_{xy}(0)/R_{xx}(0))R_{xx}(H)$. Our $R_H$ results are quantitatively consistent whether fitting to the averaged and scaled data or to the raw data, but the former approach yields better statistics. In Fig. 1b we plot $R'_{xy}(H)$ measured on SNO3 at several temperatures above and below $T_N$ (~180 K), showing linearity in $R'_{xy}(H)$ and a clear sign change in $R_H = dR'_{xy}(H)/dH$ below $T_N$.

The absolute magnitude of the resistivity for $R$NiO$_3$ can vary between bulk and thin films as reported in the literature, with thin films often being up to an order of magnitude more conductive in both the metallic and insulating phases than their bulk ceramic counterparts.[21-25] While thin films probably suffer from higher point defect concentration, bulk sintered ceramic samples suffer from a high density of grain boundaries. As noted in Ref. 8 contact resistance can be significant and may account for some of the scatter in the reported resistivity. We avoid this complication by using 4-terminal measurements for $R_{xx}$ and making Ohmic contacts to our samples by sputtered Pt metal with no intervening adhesive layer. Contact resistances were measured di-rectly using a transmission line test structure and were found to be on the order of 5 Ω at room temperature (<0.1% of the total measured resistance). Direct comparison of 4-terminal and 2-terminal resistance data recorded on the same films show that the contribution from contact resistance is consistently small throughout the full measured temperature range.

Thermopower was measured using Pt electrodes and the linear voltage technique on an as-grown samples (before etching into Hall bars).[26] The sample was suspended between two independent heaters in vacuum. At each temperature, a small thermal gradient of $\Delta T \leq 3$ K was applied to the sample, and the sample was left to reach equilibrium. The steady-state voltage difference was measured, and the Seebeck coefficient was extracted from the slope of three to five such measurements at each temperature point.

Our samples are epitaxial thin films, and appropriate care must be taken when interpreting our data as fundamental for $SmNiO_3$. Primary concerns are the effects of finite thickness and epitaxial strain. However, our films are significantly thicker than the regime (~5 unit cells) for which finite thickness strongly affects the electronic structure of the nickelates.[14,27] For 15-20 nm films, the principal effect of compressive epitaxial strain by growth on $LaAlO_3$ is to shift $T_{IM}$ downwards (see below). This is understood as the result of increasing the electronic bandwidth, which is equivalent to moving to the right in the phase diagram of Fig. 1a. Particularly relevant for this work is the quantitative similarity between our $R_H$ data on $SmNiO_3$ thin films and the results of Cheong *et al.* on bulk samples of $PrNiO_3$ and $Nd_{0.98}Sr_{0.2}NiO_3$, as we discuss below.[13] This suggests that our results for $R_H$ and interpretation in terms of the electronic structure apply to the nickelates generally.

### III. RESULTS

#### A. Structural characterization

A representative AFM micrograph from sample SNO2 is shown in Fig. 1c. Atomic step corrugation from the LaAlO$_3$ substrate is apparent, indicating that SmNiO$_3$ films grown at high sputtering pressure are smooth and continuous. The root mean square roughness of the image is ~4.5 Å. Epitaxial growth of SmNiO$_3$ is confirmed by representative $\varphi$-scans from SNO1 of the (011) pseudocubic reflection of the LaAlO$_3$ substrate and the (221) orthorhombic reflection of the SmNiO$_3$ film (Fig. 1d). The coincidence of peak angles between substrate and film shows pseudocube-on-pseudocube epitaxial growth. The lattice parameters of all SmNiO$_3$ thin films with variable oxygen stoichiometry were determined using XRD by measuring the $d$-spacing of asymmetric Bragg reflections. The four independent reflections (221), (223), (133) and (313) were measured, and the orthorhombic lattice constants were determined by a regression analysis. There is a clear trend of increasing unit cell volume with decreasing sputtering pressure due to reduced oxygen stoichiometry (Fig. 1e). Unit cell volume expansion with decreasing oxygen content is in agreement with experiments on bulk NdNiO$_{3-\delta}$ samples.[19,20]

### B. Resistivity and $T_N$

In Fig. 2 we plot the resistivity ($\rho$) as a function of temperature for our films. We show $\rho$ only for cooling and note that in previous work we did not observe appreciable thermal hysteresis (~1-2 K) between cooling and heating cycles for comparable samples.[21] We define $T_{IM}$ by the change in sign of the temperature coefficient of resistivity. For SNO1 $T_{IM} = 386 \pm 6$ K, close to the bulk value of 400 K and consistent with previous reports of SmNiO$_3$ compressively strained on LaAlO$_3$.[21,28] For the more oxygen-deficient samples SNO2 and SNO3 $T_{IM} = 380.1 \pm 0.2$ K and $375 \pm 4$ K, respectively. As oxygen vacancies are introduced, the insulating (metallic) phase becomes more (less) conductive. The same trends have been observed with Co doping in SmNi$_{1-x}$Co$_x$O$_3$,[29] Ca doping in Sm$_{1-x}$Ca$_x$NiO$_3$,[30] and oxygen reduction in NdNiO$_{3-\delta}$.[20] Therefore oxygen vacancies are probably shallow dopants in the insulating phase and scattering sites in the metallic phase. In the inset of Fig. 2 we plot $d(\ln \rho)/dT$ over a narrow temperature range. The anomalous kink in $d(\ln \rho)/dT$ is known to mark $T_N$ in $R$NiO$_3$, as directly determined by corresponding resistivity and susceptibility measurements.[8,29] In SNO1 and SNO2 we find $T_N = 214 \pm 1$ K and $211 \pm 0.1$ K, respectively, close to the bulk value of 220 K, while in SNO3 we find $T_N = 180.8 \pm 0.2$ K. The reduction of $T_N$ with oxygen vacancies is similarly consistent with the effect of doping in SmNi$_{1-x}$Co$_x$O$_3$ and Sm$_{1-x}$Ca$_x$NiO$_3$.[29,30]

### C. Transverse magnetoresistance and Seebeck coefficient

In Fig. 3a-b we present the transverse magnetoresistance (MR) for SNO2 and SNO3. We show representative data from SNO3 at select temperature points in Fig. 3a and the temperature-dependent MR at 90 kOe from both samples in Fig. 3b. There are two distinct regimes of negative MR: a broad regime of weak negative MR in the range 100 K $\leq T < T_{IM}$ (labeled "1") and a narrow regime of stronger negative MR at low temperatures $T \leq 25$ K (labeled "2"). Negative MR for $T < 20$ K has been observed in LaNiO$_3$ and was attributed to weak localization.[14] We similarly attribute the low-temperature negative MR shown here to weak localization. This is supported by the non-monotonic $\rho(H)$ (Fig. 3a) for $T \leq 25$ K, which can be explained by weak localization in the presence of spin-orbit scattering, likely due to the heavy Sm$^{3+}$ ions.[31] The broad regime of negative MR at higher temperatures is not due to weak localization, as can be seen both from the temperature scale and from the clear temperature separation between the two regimes. We propose that this instead results from the suppression of spin fluctuations in an applied magnetic field. It is known that antiferromagnetic fluctuations persist well above $T_N$ in SmNiO$_3$,[22] and that the AF order parameter does not saturate with cooling until $T < 100$ K.[32] The high temperature regime of weak, negative MR therefore coincides with the broad regime in which magnetic fluctuations are expected. This hypothesis could be tested by measurements of MR in PrNiO$_3$ and NdNiO$_3$, for which the PI phase is absent and the magnetic order parameter saturates rapidly for $T < T_N$.

We measured the Seebeck coefficient ($S$) from room temperature through the insulator-metal transition for SNO2, as shown in Fig. 3c. $S$ is negative from room temperature across $T_{IM}$, indicating electron-like majority carriers, in agreement with thermopower measurements on LaNiO$_3$,[33] PrNiO$_3$,[34] and NdNiO$_3$.[35] The abrupt jump in $S$ near $T_{IM}$ is similarly observed in PrNiO$_3$ and NdNiO$_3$ and is due to the enhanced thermopower associated with semiconductor materials with respect to metals.[34,35] The metallic phase thermopower in the aforementioned systems has an absolute value of ~10-20 µV/K, somewhat larger but within the same range (~5 µV/K) as observed here for SNO2. The thermopower temperature coefficient (d$S$/d$T$) in the metallic phase for SNO2 is -0.021 ± 0.002 µV/K$^2$, similar to that of LaNiO$_3$ (-(0.04-0.05) µV/K$^2$) and NdNiO$_3$ (-0.029 µV/K$^2$).[33,35]

### D. Hall coefficient

In Fig. 4 we present $R_H(T)$ for our three films. In the metallic phase, the sign and magnitude of $R_H$ agree well with published measurements on LaNiO$_3$,[33] PrNiO$_3$,[13] and NdNiO$_3$.[15] The most notable feature occurs with decreasing temperature as $R_H$ crosses over from hole-like at high temperature to electron-like at low temperature. This indicates that SmNiO$_3$ is a multiple-band system with both electron- and hole-like carriers, as has been shown by angle-resolved photoelectron spectroscopy (ARPES), Seebeck effect, and Hall effect measurements on LaNiO$_3$,[33,36] PrNiO$_3$,[13,34] and NdNiO$_3$.[15,35] Indeed, as shown in Fig. 3c, we also find electron-like carriers in thermopower measurements on SNO2 above room temperature. This disagreement in the sign of majority charge carriers between Hall and thermopower measurements indicates charge compen-sation. Evidence of charge compensation has also been observed in thermopower measurements of bulk NdNiO$_3$.[35]

Calculation of carrier densities and Hall mobilities in a multiple-band system is not straightforward. If we assume that there is only one hole and one electron band that contribute to the Hall effect, then the Hall coefficient at low field is given by $R_\text{H} = (p\mu_p^2 - n\mu_n^2)/e(p\mu_p + n\mu_n)^2$, where $p$ ($n$) and $\mu_p$ ($\mu_n$) are the hole (electron) densities and Hall mobilities, respectively. With only resistivity and $R_\text{H}$ data we cannot directly solve for $p$, $n$, or $\mu_{p,n}$. However, we can make additional assumptions to attempt to extract quantitative estimates of the transport parameters for the metallic phase as follows: 1) Assume $\mu_p = \mu_n = \mu$, which is reasonable given that the electron- and hole-like states are both derived from the same degenerate Ni 3d $e_g$ orbitals, and therefore their bandwidths and scattering rates should be comparable. 2) Assume $p + n = K$, a $T$-independent constant where $K = 1$ e$^-$/Ni, the free electron density appropriate for the nominal $t_{2g}^6 e_g^1$ electronic configuration of $R$NiO$_3$.[35] With these assumptions and the unit cell volumes measured directly by XRD we obtain the results in Fig. 5. We calculate the transport parameters only for temperatures in and near the metallic phase because the above assumptions are likely to break down in the insulating phase (hatched regions). We see that the hole (electron) density in the metallic phase converges towards 1.0 x 10$^{22}$ cm$^{-3}$ (0.75 x 10$^{22}$ cm$^{-3}$), independent of stoichiometry. The metallic Hall mobility for the most stoichiometric sample SNO1 is ~1 cm$^2$/V·s and the mobility decreases with increasing oxygen vacancy concentration. This is consistent with the assumption that oxygen vacancies act as scattering sites in the metallic phase, reducing the mobility. Note that quantitative analysis of $R_\text{H}$ is complicated by polaronic effects that can distort the Hall mobility relative to the drift mobility even in the metallic phase (see Section III.E.),[37] and therefore the results in Fig. 5 are only valid within the assumption that the Hall and drift mobilities are equal.

$R_H$ in Fig. 4 evolves non-monotonically with temperature and exhibits a sign crossover from hole-like to electron-like upon cooling through the insulating phase for all the SmNiO$_3$ films. We find that $R_H$ changes sign at 197.8 ± 0.4 K (SNO1), 184 ± 4 K (SNO2), and 117 ± 4 K (SNO3). This crossover occurs somewhat below $T_N$ and the crossover temperatures track the evolution of $T_N$ with oxygen stoichiometry. This implies that the change in majority carrier type with temperature may be associated with a change in electronic structure due to antiferromagnetic ordering. The connection between the Néel transition and the sign change of $R_H$ in our SmNiO$_3$ thin films is supported by data on bulk ceramic samples of PrNiO$_3$ and Nd$_{0.98}$Sr$_{0.2}$NiO$_3$ published by Cheong *et al.*[13] We reproduce their results for PrNiO$_3$ in Fig. 4 (inset); their results for Nd$_{0.98}$Sr$_{0.2}$NiO$_3$ are quantitatively similar. For both PrNiO$_3$ and Nd$_{0.98}$Sr$_{0.2}$NiO$_3$, $R_H$ changes sign near $T_{IM}$ and for these materials $T_N = T_{IM}$. The quantitative agreement between our results on SmNiO$_3$, for which $T_N < T_{IM}$, and those of Cheong *et al.* on materials for which $T_N = T_{IM}$ illustrates that the crossover in $R_H$ is not specific to thin films and it suggests an underlying cause common to the Néel transition in $R$NiO$_3$. We discuss below a possible mechanism for the AF order in SmNiO$_3$ thin films based on the electronic bandstructure that can explain both the unusual antiferromagnetic wavevector and the link between $T_N$ and $R_H$, including the observation that the crossover in $R_H$ occurs somewhat below $T_N$.

$R_H$ appears to tend toward zero as $T \rightarrow 0$ K, suggesting nearly perfect charge compensation at very low temperature. Similar behavior has been observed in the AF phase of the correlated electron insulator Na$_{0.5}$CoO$_2$,[38] and it has been ascribed to formation of doubly- and singly-occupied sublattices in the ground state, where the singly-occupied sublattice is antiferromagnetic and the ground state has particle-hole symmetry.[17]

**E. Small polarons, hopping conductivity, and the Hall coefficient**

Quantitative interpretation of $R_H$ is complicated by the presence of small polarons in $R$NiO$_3$. In particular, the Hall mobility $\mu_H$ may differ significantly from the drift mobility for $T > T_t$, where $T_t \sim (1/2)\Omega/k_B$ and $\Omega$ is a characteristic optical phonon frequency.[37,39] For polaronic materials such as $R$NiO$_3$ $T_t$ is the temperature above which electronic bandstructure is no longer a valid concept because rapid phonon-assisted transitions smear out individual states in reciprocal space.[39] For $R$NiO$_3$ we estimate $T_t \sim 450$ K, using for $\Omega$ the frequency 600 cm$^{-1}$ of the NiO$_6$ octahedra breathing mode that is thought to be most responsible for the formation of small polarons.[40] For $T > T_t$ charge transport is expected to occur by diffusive hopping, and $\mu_H$ is controlled by the connectivity and dimensionality of the lattice. For $R$NiO$_3$ this means that $\mu_H$ for $T > T_t$ will depend strongly on $t'/t$, where $t$ and $t'$ are the nearest- and next-nearest neighbor hopping integrals that describe hopping between Ni sites along a square edge and a square diagonal, respectively. This would be an interesting direction for future study but may be complicated by the thermodynamic instability of $R$NiO$_3$ at high temperatures.

For temperatures $T < T_t \sim 450$ K the effect of small polarons on $R_H$ is much less pronounced, and $\mu_H$ is close to the standard drift mobility of a band insulator or conductor.[37] In particular the *sign* of $\mu_H$ corresponds to the sign of the charge carriers, and therefore our interpretation of the crossover of $R_H$ below $T_N$ remains valid. However, the magnitude of $R_H$ still depends on the ratio $t'/t$. This dependence complicates any quantitative analysis of $R_H$, and may affect the precise temperature of the sign crossover.

The dominant transport mechanism for $T < T_{IM}$ is not well understood. $\rho(T<T_{IM})$ for the nickelates is not well modeled by a single mechanism and is likely due to an interplay of thermal activation and hopping in the presence of disorder.[41] Our data in Fig. 2 is consistent with this picture and is not well described by a model of thermal activation or variable range hopping over any appreciable temperature range, although for temperatures below 100 K there is evidence (not shown) that mobility is controlled by variable range hopping in the presence of Coulomb interactions. As with the above case of diffusive transport of small polarons $R_H$ can be affected by the hopping mechanism, and therefore the precise temperature of the sign crossover in $R_H$ may be affected. It is possible that the vanishing of $R_H$ as $T \rightarrow 0$ K may be connected to emergence of variable range hopping as the mobility limiting mechanism at low temperatures. However without a complete understanding of the transport mechanism, and in particular the ratio $t'/t$, it is challenging to estimate this effect.

## IV. DISCUSSION

### A. Nickelate magnetism and the case of $T_N = T_{IM}$

Antiferromagnetism in $R$NiO$_3$ develops at wavevector $q = (½, 0, ½)$, unique among the perovskite-derived complex oxides, and it cannot be described by a set of local magnetic interactions with the symmetry of the orthorhombic crystal lattice.[32] An explanation requires either some local order that yields site-dependent magnetic interactions or a non-local mechanism. The nominal electronic configuration $t_{2g}^6 e_g^1$ of Ni$^{3+}$ in $R$NiO$_3$ is orbitally degenerate, which led to early proposals that collective Jahn-Teller orbital order was responsible for the antiferromagnetism.[32,42] This possibility has since been ruled out by symmetry: the symmetry required for a description involving orbital order is $Bb2_1m$ but the symmetry in the AFI phase is $P2_1/n$.[32,43] Furthermore, experiments have searched for and have not found orbital order using resonant X-ray diffraction.[43,44] Adjacent $NiO_6$ octahedra do differentiate by bond length below $T_{IM}$,[45] but whether this is the result of charge transfer or a heterogeneous assortment of covalent-like and ionic-like Ni-O bonds remains unclear.[10,44,46,47] A recent study using dynamical mean field theory found differentiation into covalent-like and ionic-like octahedra without charge transfer but incorrectly predicts a ferromagnetic ground state,[10] thus suggesting that non-local interactions are needed to properly describe the antiferromagnetism.

The necessary non-local mechanism could come from the bandstructure. In Fig. 6a we present the Fermi surface of metallic $LaNiO_3$ calculated with a two-band tight-binding model by Lee et al.[9] The Fermi surface features a large, well nested hole-like cube (blue hatching) centered at the R point and a small electron-like pocket (red hatching) centered at the Γ point. Additional bandstructure calculations using density functional theory[48] and measurements on $LaNiO_3$ using ARPES[36] are in good agreement with Lee et al. for metallic $RNiO_3$. For this bandstructure the spin susceptibility peaks at a nesting vector (Fig. 6a) equal to the experimentally measured AF wavevector $q$.[9] The bandstructure therefore provides a natural explanation for the observed antiferromagnetic order in terms of Fermi surface nesting, and it correctly predicts the signs of the Hall and Seebeck coefficients in the metallic phase.[9]

The bandstructure can also explain the experimentally observed crossover in $R_H$ from hole-like to electron-like occurring at or below $T_N$. Calculations show that only the hole-like surface is nested by $q$ and is gapped by the onset of antiferromagnetism. The Néel transition therefore affects charge compensation by preferentially moving hole-like states farther from the Fermi energy, and the crossover in $R_H$ occurs when a sufficient fraction of the hole-like states have been removed from the population of mobile carriers. A crossover in the sign of $R_H$ due to density wave formation is an understood phenomenon in electronic conductors; examples include

NbSe$_2$,[49] α-U,[50] and tungsten bronze.[51] The temperature of the crossover depends on the precise details of the bandstructure and the evolution of the magnetic order parameter. For perfectly nested hole-like surfaces the crossover should closely coincide with $T_N$. For a realistic Fermi surface the crossover should occur somewhat below $T_N$, as an increasing fraction of hole-like states are gapped by the increasing magnetic order parameter.

Therefore, as has been noted previously,[9,52] antiferromagnetism in those materials (PrNiO$_3$, NdNiO$_3$) for which $T_N = T_{IM}$ is consistent with the SDW mechanism. The nested hole-like Fermi surface justifies the otherwise unexplained wavevector *q*. The signs of *S* and $R_H$, including the crossover in $R_H$ below $T_N$, are also explained by the calculated bandstructure. Fermi surface nesting alone does not determine the magnetic energy scale, which is required to understand the magnitude of $T_N$ and the saturated ordered magnetic moment. In a material with strong electron-electron and electron-phonon interactions the magnetic energy scale is enhanced relative to the case of an SDW in a non-interacting material (see below). For *R*NiO$_3$, the energy scale is likely determined by the superexchange mechanism.[8]

### B. Bandstructure and antiferromagnetism in the case of $T_N < T_{IM}$

The concept of SDW antiferromagnetism is conventionally expected to be inapplicable to those materials (*e.g.* SmNiO$_3$) for which $T_N < T_{IM}$ and antiferromagnetic order develops from an insulator. However, SDW antiferromagnetism may yet apply to materials which are insulating, but for which electronic bandstructure remains a valid concept (*i.e.* a "collective electron" insulator with *b* larger than but nearly equal to $b_c$, in the language of Goodenough,[11] where *b* is a measure of interactions between neighboring d-electrons and $b_c$ is the minimum value of *b* for which band theory is applicable). For such materials crystal momentum is a good quantum num-ber, and it is meaningful to consider how the occupied single particle energy levels are affected by a SDW. Here we discuss why the SDW mechanism might apply to SmNiO$_3$.

The nickelates are frequently classified as charge transfer insulators with respect to the Zaanen-Sawatzky-Allen scheme for correlated-electron oxides.[53] This means that in the insulating phase the valence band is derived from oxygen 2p orbitals ($3d^7 2p^6$ configuration), the conduction band is the upper Hubbard band derived from correlated Ni 3d $e_g$ orbitals ($3d^8\underline{L}$ configuration, $\underline{L}$ a ligand hole), and the bandgap depends on the energy ($\Delta$) of the transition $3d^7 2p^6 \rightarrow 3d^8\underline{L}$. However, based on the results of photoemission and x-ray absorption spectroscopy it is known that the valence band of $R$NiO$_3$ has significant $e_g$ character and that the electronic ground state is a mixture of $3d^7 2p^6$ and $3d^8\underline{L}$.[54,55] For example, based on photoelectron spectra and a configuration interaction calculation Mizokawa *et al.* found that the valence band of PrNiO$_3$ in the AFI phase is majority $3d^8\underline{L}$ in character.[54] Moreover, based on an analysis of x-ray absorption Medarde *et al.* found a significant $3d^8\underline{L}$ contribution to the valence band of PrNiO$_3$ and NdNiO$_3$, and that the mixing between $3d^7 2p^6$ and $3d^8\underline{L}$ changes little across $T_{IM}$.[55] The same authors suggested that $R$NiO$_3$ might be a negative-$\Delta$ charge transfer insulator if not for thermopower results showing negative charge carriers; Hall results showing positive carriers were not available at that time. No matter whether $\Delta$ is small and positive, or small and negative, $R$NiO$_3$ is a covalent insulator with significant $e_g$ orbital contribution to the occupied valence band in the insulating phase.

As discussed in Ref. 9 the low energy (near $E_F$) electronic features of the PM phase are captured by the bandstructure arising from one electron in the $e_g$ manifold, and the details do not depend strongly on the precise covalency. Therefore Fermi surface nesting is robust for metallic $R$NiO$_3$, and the $e_g$ character of the valence band for $T < T_{IM}$ implies that states near the nested

Fermi surface for $T > T_{IM}$ remain occupied for $T < T_{IM}$. The question then becomes: How do the electronic energy levels change upon entering the PI phase?

The origin of the insulating phase in the nickelates remains poorly understood. We do not ascribe a specific mechanism here, but regardless, we emphasize that crystal momentum remains a good quantum number (*i.e.* electronic bandstructure likely remains a valid concept) for $T < T_{IM}$. The alternative is that the insulating phase is characterized by diffusive motion of small polarons.[39] However, for polaronic materials such as $R$NiO$_3$ electronic bandstructure is a valid concept at low temperature and breaks down above temperature $T_t \sim (1/2)\Omega/k_B \sim 450$ K (see Section III.E). It is therefore apparent that the PI phase of $R$NiO$_3$ is a "collective electron" insulator as described by Goodenough,[11] for which crystal momentum remains a good quantum number and it is meaningful to construct a band diagram for single quasiparticle energy levels.

Without ascribing a particular mechanism to the insulator-metal transition, we can still describe the change in electronic structure at the nested Fermi surface upon cooling through $T_{IM}$. In Fig. 6b we illustrate the effect of opening a bandgap at the Fermi surface due to unit cell doubling: the occupied states are lowered in energy, and the unoccupied states are raised.[11] This generic mechanism would apply to SmNiO$_3$ in most of the scenarios proposed in the literature to explain the insulator-metal transition, including charge disproportionation, orbital order such as a collective Jahn-Teller distortion, or bond order driven by electron correlation energy such as a site-selective Mott transition.[25,43,45,52,56] The salient point for this study is that for small values of the bandgap the valence band can remain somewhat well nested even for $T < T_{IM}$, and this allows for further lowering of the net system energy by SDW formation.

The energy lowering possible through SDW formation at $T_N < T_{IM}$ depends on the relative values of the insulating bandgap ($E_g$) and the SDW spin-flip energy ($2\Delta_{SDW}$); for $2\Delta_{SDW} \sim E_g$ or $2\Delta_{SDW} > E_g$ the energy lowering may be appreciable. $E_g$ in $R$NiO$_3$ is not well known, but it is small and temperature-dependent. Electrical measurements on NdNiO$_3$ suggest $E_g \sim 50$ meV,[35,41] and if we fit our $\rho(T)$ data to an activated form we find $E_g \sim 100$ meV for temperatures near $T_N$. We can estimate $2\Delta_{SDW}$ from the expression $2\Delta_{SDW} = ck_BT_N$. For weakly coupled systems the prefactor $c = 3.5$, but it is enhanced in the presence of strong electron correlation and electron-phonon coupling.[57] Here we use $c = 10$,[58] an approximate lower bound determined from studies of charge order in La$_{1-x}$Ca$_x$MnO$_3$. With $T_N = 220$ K we find $2\Delta_{SDW} \sim 200$ meV. These conservative estimates give $2\Delta_{SDW}/E_g \sim 2$. The ordering of energy scales $2\Delta_{SDW} \geq E_g$ is notable given that $T_N < T_{IM}$. However, it is consistent with the suppression of the critical temperature (represented by the factor $c$) for itinerant density waves in materials with strongly coupled degrees of freedom.

The net energy lowering due to SDW formation at $T_N < T_{IM}$ is illustrated in Fig. 6c, where we plot the band dispersion for the case $2\Delta_{SDW}/E_g = 2$. Due to the pre-existing insulating gap the energy of states at $k = ¼$ is not lowered by the full $\Delta_{SDW}$. However, the net energy lowering can be large if the bands remain somewhat well nested. The SDW affects most strongly those states for which $|E(k\pm q)-E(k)| \leq \Delta_{SDW}$ and therefore the net energy lowering and the amplitude of the SDW is further enhanced if the bandwidth is small, as expected for a polaronic material. For the illustration in Fig. 6 we have use the estimate $2\Delta_{SDW}/E_g = 2$, but this mechanism should remain valid as long as $2\Delta_{SDW}$ is comparable in magnitude to $E_g$.

It is remarkable to see that the influence of the bandstructure may extend deep within the PI phase of SmNiO$_3$ where the Fermi surface no longer exists. This concept of a density wave driven by bandstructure, but not necessarily the Fermi surface *per se*, has emerged recently in theoretical treatments of systems that fall between the limits of weak and strong electronic correla-tion. SDW magnetism in the iron pnictides is stabilized by bandstructure, but the Fermi surface may not play a determinative role.[18] Another example is the correlated covalent insulator $Na_{0.5}CoO_2$, which shares with $SmNiO_3$ an AFI ground state, a crossover in $R_H$ near $T_N$, and a non-monotonic $R_H(T)$ with $R_H \to 0$ as $T \to 0$.[17] In $Na_{0.5}CoO_2$ an SDW is responsible for a crossover in the sign of $R_H$ at $T_N$,[16] but both the exchange and electron correlation interactions must be considered to describe the low-temperature transport and the non-monotonic $R_H(T)$. Likewise, for $R$NiO$_3$ the SDW is not responsible for the insulating state, but both the SDW mechanism and the strong correlations must be considered to describe the AFI state.

### C. Connections between band and superexchange magnetism

For $R$ heavier than Sm the dependence of $T_N$ on bandwidth is in qualitative agreement with the theory of superexchange magnetism.[8] This agreement starts breaking down for Sm, although it is unclear whether this represents a true breakdown of the superexchange theory or a more complicated dependence of the parameters on $r(R^{3+})$ combined with greater difficulty in interpreting the experimental data.[8] In either case, the antiferromagnetic energy scale is largely determined by the superexchange mechanism. However, non-local interactions remain essential to understanding the observed antiferromagnetic order. $\boldsymbol{q} = (½, 0, ½)$ can in fact be derived from superexchange interactions in the strong coupling limit but only by considering both nearest- and next-nearest-neighbor hopping.[9] The importance of non-zero next-nearest-neighbor hopping supports our claim that bandstructure is a valid concept in the insulating phases of $R$NiO$_3$.

### V. CONCLUSIONS

We present resistivity, magnetoresistance, Seebeck coefficient, and Hall coefficient measurements of epitaxial SmNiO$_3$ thin films with varying oxygen content. The Hall coefficient measurements span a wide (30 – 400 K) temperature range through both Néel and insulator-metal phase transitions. We observe a hole-like Hall coefficient and electron-like Seebeck coefficient from room temperature through the insulator-metal transition. By varying the oxygen stoichiometry of our films we show that the Néel transition induces a crossover in the sign of the Hall coefficient from hole- to electron-like. We suggest that valence band states in the insulating phase support the observed antiferromagnetic order via a mechanism akin to SDW magnetism in metallic systems. Electronic structure calculations in the PI phase will be necessary to confirm the proposed model but at present are challenging due to uncertainty over the nature of the insulating state. We also note that the magnetic phase diagram of $R$NiO$_3$ looks like that of a band of correlated electrons at half-filling instead of quarter-filling.[11] This observation may be useful in ongoing efforts to understand the bandwidth-controlled insulator-metal transition in $R$NiO$_3$.

The case of SmNiO$_3$ illustrates connections between concepts developed to treat the limits of electron localization (oxide physics, superexchange magnetism) and delocalization (metals physics, band magnetism). SmNiO$_3$ marks the point in the nickelate phase diagram that is most challenging to describe within the conceptual framework of either weak or strong coupling.[11,12] Therefore it is perhaps unsurprising to see that bandstructure plays an important role in the insulating phase, or that the superexchange interaction is the dominant magnetic energy scale in a system with significant next-nearest-neighbor hopping. These connections suggest a flexible conceptual framework with which to understand the unusual antiferromagnetic order in the nickelates.

**Acknowledgements**

The authors acknowledge the ARO MURI (W911-NF-09-1-0398), NSF (DMR-0952794), and AFOSR (FA9550-12-1-0189) for financial support. This work was performed in part at the Center for Nanoscale Systems at Harvard University, which is supported under NSF award ECS-0335765, and shared facilities of the University of Chicago MRSEC. The authors acknowledge important discussions with and contributions from SungBin Lee and Leon Balents. R. J. acknowledges helpful discussions with Jasper van Wezel and Ramona Levine.

**References**

1     J. C. Loudon, S. Cox, A. J. Williams, J. P. Attfield, P. B. Littlewood, P. A. Midgley, and N. D. Mathur, Phys. Rev. Lett. **94**, 097202 (2005).

2     M. A. Mroginski, N. E. Massa, H. Salva, J. A. Alonso, and M. J. Martínez-Lope, Phys. Rev. B **60**, 5304 (1999).

3     M. K. Stewart, J. Liu, M. Kareev, J. Chakhalian, and D. N. Basov, Phys. Rev. Lett. **107**, 176401 (2011).

4     M. Gibert, P. Zubko, R. Scherwitzl, J. Íñiguez, and J.-M. Triscone, Nat. Mater. **11**, 195 (2012).

5     D. G. Ouellette, S. Lee, J. Son, S. Stemmer, L. Balents, A. J. Millis, and S. J. Allen, Phys. Rev. B **82**, 165112 (2010).

6     M. L. Medarde, J. Phys.: Condens. Matter **9**, 1679 (1997).

7     G. Catalan, Phase Transitions **81**, 729 (2008).

8     J. S. Zhou, J. B. Goodenough, and B. Dabrowski, Phys. Rev. Lett. **95**, 127204 (2005).

9     S. Lee, R. Chen, and L. Balents, Phys. Rev. B **84**, 165119 (2011).

10    H. Park, A. J. Millis, and C. A. Marianetti, Phys. Rev. Lett. **109**, 156402 (2012).

11    J. B. Goodenough, Prog. Solid State Chem. **5**, 145 (1971).

12    J. E. Hirsch, Phys. Rev. B **35**, 1851 (1987).

13    S. W. Cheong, H. Y. Hwang, B. Batlogg, A. S. Cooper, and P. C. Canfield, Physica B **194-196**, 1087 (1994).

14    J. Son, P. Moetakef, J. M. LeBeau, D. Ouellette, L. Balents, S. J. Allen, and S. Stemmer, Appl. Phys. Lett. **96**, 062114 (2010).

15   R. Scherwitzl, P. Zubko, I. G. Lezama, S. Ono, A. F. Morpurgo, G. Catalan, and J.-M. Triscone, Adv. Mater. **22**, 5517 (2010).

16   J. Bobroff, G. Lang, H. Alloul, N. Blanchard, and G. Collin, Phys. Rev. Lett. **96**, 107201 (2006).

17   T.-P. Choy, D. Galanakis, and P. Phillips, Phys. Rev. B **75**, 073103 (2007).

18   M. D. Johannes and I. I. Mazin, Phys. Rev. B **79**, 220510 (2009).

19   A. Tiwari and K. P. Rajeev, Solid State Commun. **109**, 119 (1999).

20   I. V. Nikulin, M. A. Novojilov, A. R. Kaul, S. N. Mudretsova, and S. V. Kondrashov, Mater. Res. Bull. **39**, 775 (2004).

21   S. D. Ha, M. Otaki, R. Jaramillo, A. Podpirka, and S. Ramanathan, J. Solid State Chem. **190**, 233 (2012).

22   J. Pérez-Cacho, J. Blasco, J. García, M. Castro, and J. Stankiewicz, J. Phys.: Condens. Matter **11**, 405 (1999).

23   M. T. Escote, A. M. L. da Silva, J. R. Matos, and R. F. Jardim, J. Solid State Chem. **151**, 298 (2000).

24   F. Conchon, et al., J. Appl. Phys. **103**, 123501 (2008).

25   J. S. Zhou, J. B. Goodenough, and B. Dabrowski, Phys. Rev. B **67**, 020404 (2003).

26   J. Martin, T. Tritt, and C. Uher, J. Appl. Phys. **108**, 121101 (2010).

27   R. Scherwitzl, S. Gariglio, M. Gabay, P. Zubko, M. Gibert, and J. M. Triscone, Phys. Rev. Lett. **106**, 246403 (2011).

28   F. Conchon, A. Boulle, R. Guinebretiere, C. Girardot, S. Pignard, J. Kreisel, F. Weiss, E. Dooryhee, and J.-L. Hodeau, Appl. Phys. Lett. **91**, 192110 (2007).

29   J. Pérez-Cacho, J. Blasco, J. García, and J. Stankiewicz, Phys. Rev. B **59**, 14424 (1999).

| 30 | P. H. Xiang, S. Asanuma, H. Yamada, I. H. Inoue, H. Akoh, and A. Sawa, Appl. Phys. Lett. **97**, 032114 (2010). |
|---|---|
| 31 | G. Bergmann, Phys. Rep. **107**, 1 (1984). |
| 32 | J. Rodríguez-Carvajal, S. Rosenkranz, M. Medarde, P. Lacorre, M. T. Fernandez-Díaz, F. Fauth, and V. Trounov, Phys. Rev. B **57**, 456 (1998). |
| 33 | N. Gayathri, A. K. Raychaudhuri, X. Q. Xu, J. L. Peng, and R. L. Greene, J. Phys.: Condens. Matter **10**, 1323 (1998). |
| 34 | X. Granados, J. Fontcuberta, X. Obradors, and J. B. Torrance, Phys. Rev. B **46**, 15683 (1992). |
| 35 | X. Granados, J. Fontcuberta, X. Obradors, L. Mañosa, and J. B. Torrance, Phys. Rev. B **48**, 11666 (1993). |
| 36 | R. Eguchi, A. Chainani, M. Taguchi, M. Matsunami, Y. Ishida, K. Horiba, Y. Senba, H. Ohashi, and S. Shin, Phys. Rev. B **79**, 115122 (2009). |
| 37 | L. Friedman and T. Holstein, Ann. Phys. **21**, 494 (1963). |
| 38 | M. L. Foo, Y. Wang, S. Watauchi, H. W. Zandbergen, T. He, R. J. Cava, and N. P. Ong, Phys. Rev. Lett. **92**, 247001 (2004). |
| 39 | T. Holstein, Ann. Phys. **8**, 343 (1959). |
| 40 | F. P. de la Cruz, C. Piamonteze, N. E. Massa, H. Salva, J. A. Alonso, M. J. Martínez-Lope, and M. T. Casais, Phys. Rev. B **66**, 153104 (2002). |
| 41 | G. Catalan, R. M. Bowman, and J. M. Gregg, Phys. Rev. B **62**, 7892 (2000). |
| 42 | J. L. García-Muñoz, J. Rodríguez-Carvajal, and P. Lacorre, Europhys. Lett. **20**, 241 (1992). |

43   V. Scagnoli, U. Staub, A. M. Mulders, M. Janousch, G. I. Meijer, G. Hammerl, J. M. Tonnerre, and N. Stojic, Phys. Rev. B **73**, 100409 (2006).

44   U. Staub, G. I. Meijer, F. Fauth, R. Allenspach, J. G. Bednorz, J. Karpinski, S. M. Kazakov, L. Paolasini, and F. d'Acapito, Phys. Rev. Lett. **88**, 126402 (2002).

45   J. A. Alonso, M. J. Martínez-Lope, M. T. Casais, M. A. G. Aranda, and M. T. Fernández-Díaz, J. Am. Chem. Soc. **121**, 4754 (1999).

46   J. B. Goodenough, J. S. Zhou, F. Rivadulla, and E. Winkler, J. Solid State Chem. **175**, 116 (2003).

47   M. Medarde, C. Dallera, M. Grioni, B. Delley, F. Vernay, J. Mesot, M. Sikora, J. A. Alonso, and M. J. Martínez-Lope, Phys. Rev. B **80**, 245105 (2009).

48   N. Hamada, J. Phys. Chem. Solids **54**, 1157 (1993).

49   H. N. S. Lee, H. McKinzie, D. S. Tannhauser, and A. Wold, J. Appl. Phys. **40**, 602 (1969).

50   G. M. Schmiedeshoff, et al., Philos. Mag. **84**, 2001 (2004).

51   C. Hess, C. Schlenker, J. Dumas, M. Greenblatt, and Z. S. Teweldemedhin, Phys. Rev. B **54**, 4581 (1996).

52   S. Lee, R. Chen, and L. Balents, Phys. Rev. Lett. **106**, 016405 (2011).

53   J. Zaanen, G. A. Sawatzky, and J. W. Allen, Phys. Rev. Lett. **55**, 418 (1985).

54   T. Mizokawa, A. Fujimori, T. Arima, Y. Tokura, N. Mōri, and J. Akimitsu, Phys. Rev. B **52**, 13865 (1995).

55   M. Medarde, A. Fontaine, J. L. García-Muñoz, J. Rodríguez-Carvajal, M. de Santis, M. Sacchi, G. Rossi, and P. Lacorre, Phys. Rev. B **46**, 14975 (1992).

[56] I. I. Mazin, D. I. Khomskii, R. Lengsdorf, J. A. Alonso, W. G. Marshall, R. M. Ibberson, A. Podlesnyak, M. J. Martínez-Lope, and M. M. Abd-Elmeguid, Phys. Rev. Lett. **98**, 176406 (2007).

[57] R. Jaramillo, Y. Feng, J. C. Lang, Z. Islam, G. Srajer, H. M. Rønnow, P. B. Littlewood, and T. F. Rosenbaum, Phys. Rev. B **77**, 184418 (2008).

[58] K. H. Kim, S. Lee, T. W. Noh, and S. W. Cheong, Phys. Rev. Lett. **88**, 167204 (2002).

**Figure 1:** (a) Phase diagram of $R$NiO$_3$. See Refs. 6, 7, and 8 for selected source data. (b) Hall resistance $R'_{xy}$ at select temperatures for SNO3, showing linearity and crossover in the sign of $R_\text{H}$ below $T_\text{N} \sim 180$ K. (c) AFM micrograph of SNO2 showing atomic steps (0.379 nm) of LaAlO$_3$ (001) substrate. (d) XRD $\varphi$-scans from SNO1 at (011) pseudocubic reflection of LaAlO$_3$ and (221) orthorhombic reflection of SmNiO$_3$. (e) Unit cell volume of SmNiO$_3$ grown on LaAlO$_3$ as a function of total sputtering pressure.

**Figure 2:** $\rho(T)$ for SNO1, SNO2, and SNO3 showing the effects of varying oxygen stoichiometry. Oxygen content is monotonically decreased from SNO1 to SNO3. Inset: d(ln $\rho$)/d$T$, where the anomalous kink marks $T_\text{N}$.

**Figure 3:** (a) Transverse magnetoresistance $MR(H) \equiv (\rho(H) - \rho(0))/\rho(0)$ at select temperatures for SNO3. Data are collected in full ±90 kOe field sweeps; the component even in $H$ is extracted and plotted here. (b) $MR$(90 kOe) as a function of temperature for SNO2 and SNO3. $MR < 0$ and $MR > 0$ data are shown on separate logarithmic plots. (1) and (2) denote distinct regimes of negative MR. (c) Seebeck coefficient measured on SNO2 showing electron-like majority carriers in insulating and metallic phases above room temperature.

**Figure 4:** $R_\text{H}(T)$ for SNO1, SNO2, and SNO3. The crossover in sign tracks the evolution of $T_\text{N}$ with oxygen stoichiometry. Inset: $R_\text{H}(T)$ and $\rho(T)$ for PrNiO$_3$ with $T_\text{IM} = T_\text{N}$ marked. Reprinted from Ref. 13, Copyright 1994, with permission from Elsevier.

**Figure 5:** (a) Estimated hole and electron carrier densities and (b) Hall mobilities for SNO1, SNO2, and SNO3 in and near the metallic regime extracted from Hall coefficient data. Transport data are calculated assuming $\mu_p = \mu_n = \mu$ and $p + n = K$, a constant. The hatched area represents the likely breakdown of the assumptions for $T < T_{IM}$ (see text).

**Figure 6:** (a) Fermi surface of LaNiO$_3$. The hole-like cube (blue) is shown in the left panel; the right panel shows both the electron-like pocket (red) and the hole-like surfaces in a side view projection, with the calculated nesting vector illustrated. Adapted with permission from Ref. 9. Copyrighted by the American Physical Society. Short dashed line is cut along which dispersion is plotted in lower panel. (b-c) Schematic of dispersion along a short line crossing the nested Fermi surface as the temperature is lowered sequentially through $T_{IM}$ and $T_N$. (b) Dispersion for $T > T_{IM}$ (solid line) and $T_{IM} > T > T_N$ (dotted line) for a generic insulating bandgap. Occupied states are lowered in energy by $E_g/2$, as illustrated by the grey filling, and unoccupied states are raised. (c) Dispersion for $T_{IM} > T > T_N$ (dotted line) and $T_N > T$ (dashed line) for $2\Delta_{SDW}/E_g = 2$. The energy of the occupied states at $k = ¼$ is not lowered by the full $\Delta_{SDW}$, but the overall energy lowering (grey filling) may be significant.

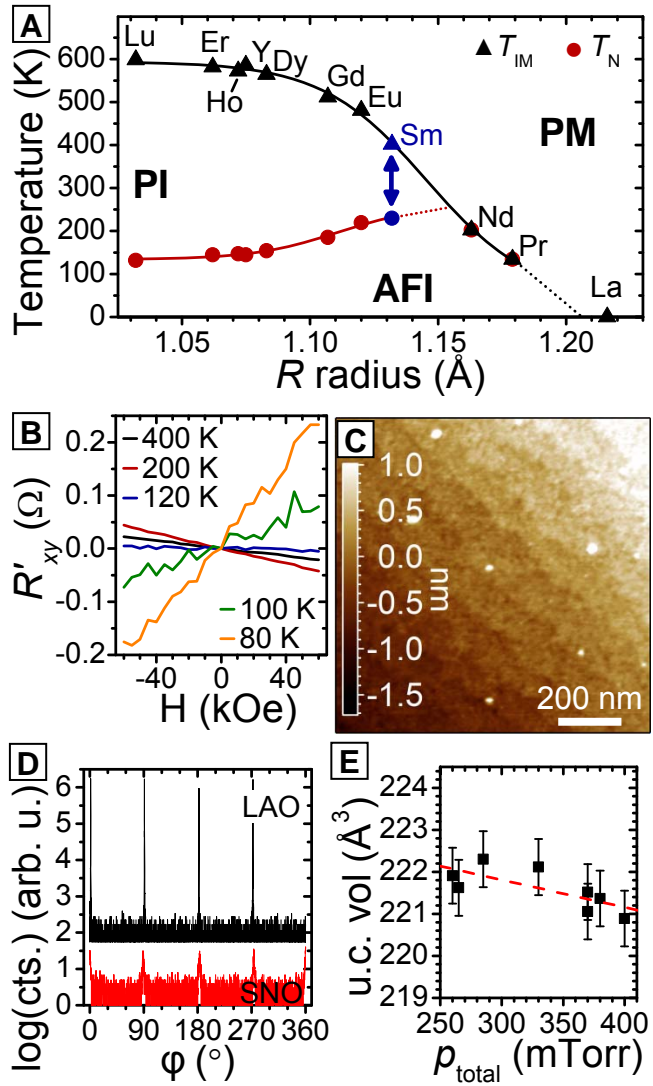

Figure 1

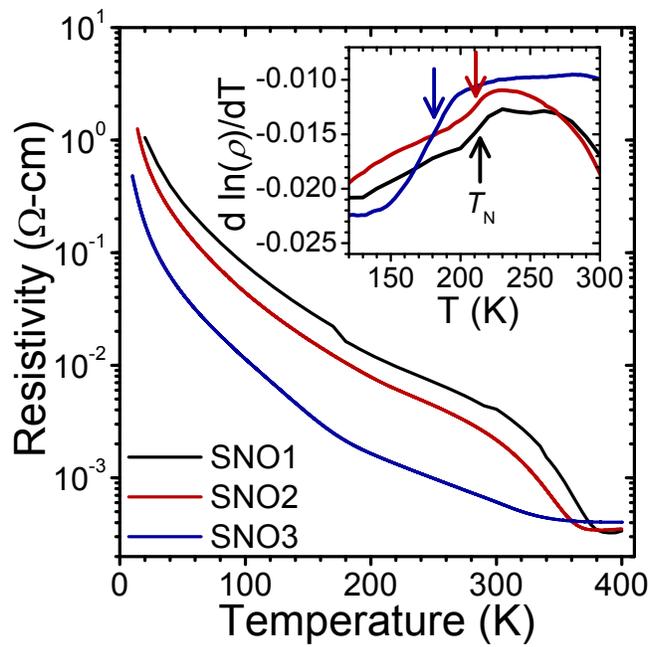

Figure 2

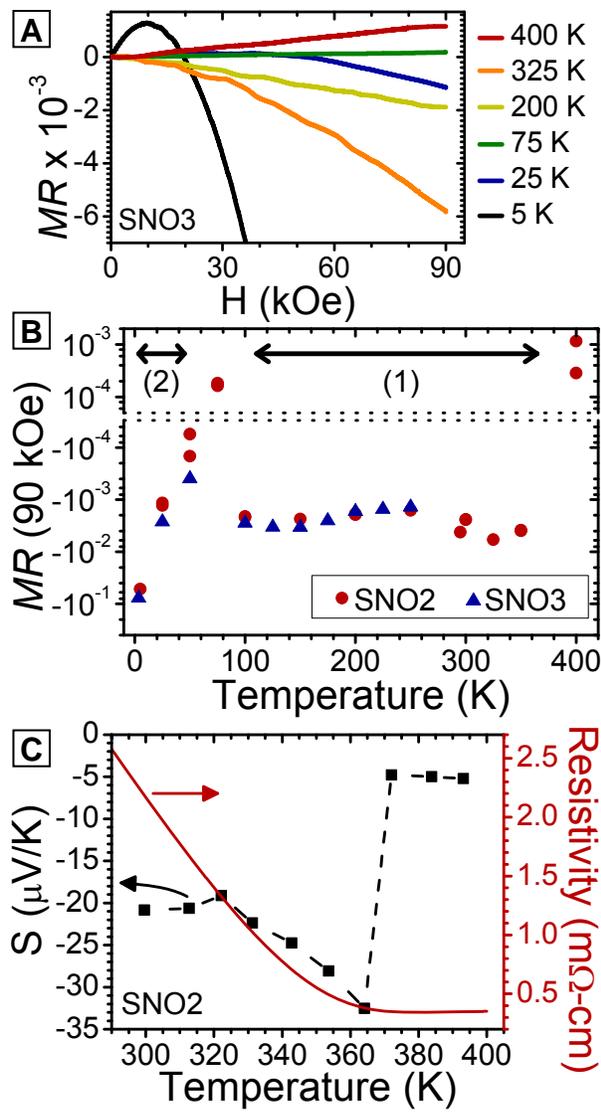

Figure 3

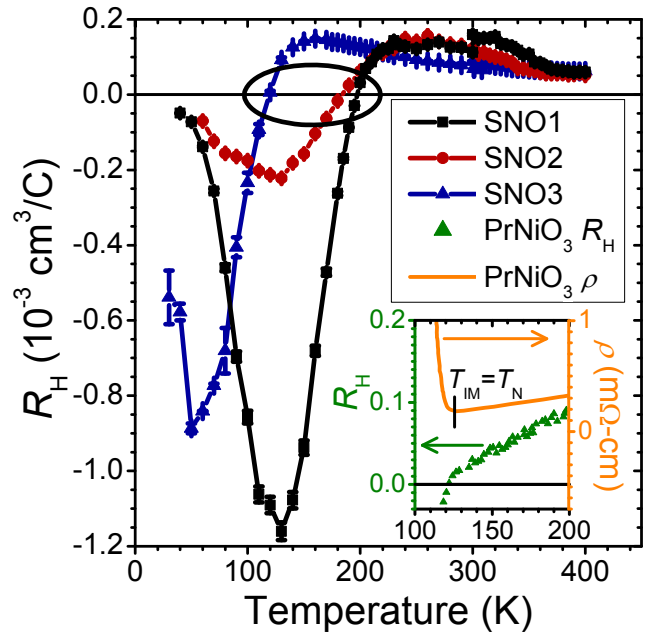

Figure 4

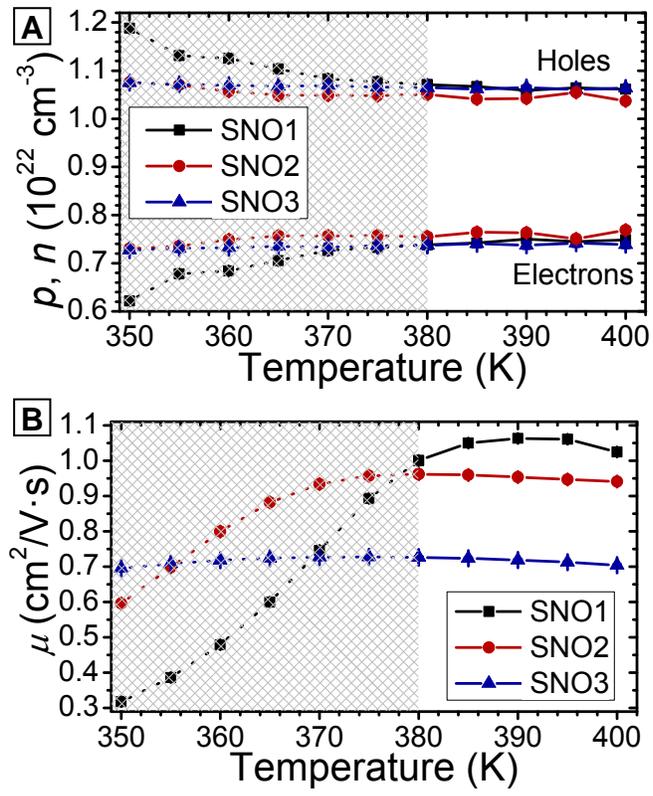

Figure 5

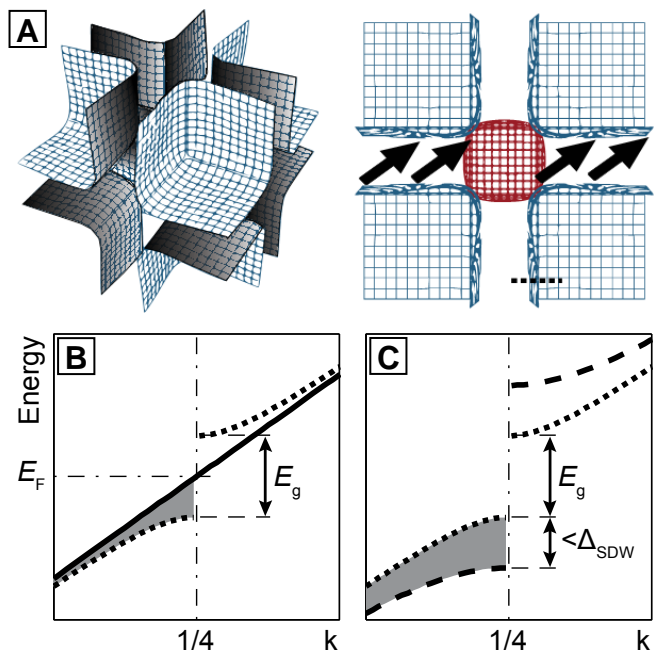

Figure 6